\documentclass[final,10pt,a4paper]{IEEEtran}

\usepackage[latin1]{inputenc}
\usepackage{cite}
\usepackage[pdftex]{graphicx}
 \graphicspath{{../pdf/}{../jpeg/}}
 \DeclareGraphicsExtensions{.pdf,.jpeg,.png}
\usepackage{pgf}
\usepackage{pgfplots}
\usepackage[cmex10]{amsmath}
\usepackage{amssymb}
\usepackage{array}
\usepackage{mdwmath}
\usepackage{mdwtab}
\usepackage{url}
\usepackage{fixedmultirow}
\usepackage{trfsigns}
\usepackage{fancyhdr}
\usepackage{nicefrac}
\usepackage{mathtools} 
\usepackage{algorithm}
\usepackage{algpseudocode}
\algrenewcommand\algorithmicindent{.5em}

\graphicspath{{graphics/}}
\graphicspath{{../Bilder/}}

\usetikzlibrary{calc,trees,positioning,arrows,chains,shapes.geometric,%
 decorations.pathreplacing,decorations.pathmorphing,shapes,%
 matrix,shapes.symbols,plotmarks,decorations.markings,shadows,fit,patterns}
\usetikzlibrary{spy,backgrounds}
\usepgfplotslibrary{groupplots}

\tikzset{every picture/.style={>=latex}} 
\pgfplotsset{compat=1.3}
\pgfplotsset{filter discard warning=false} 
\pgfplotsset{every axis label/.append style={font=\small}}
\pgfplotsset{every tick label/.append style={font=\footnotesize}}
\pgfsetplotmarksize{2pt}

\tikzstyle{EMTY}       =  [ fill=white, ]
\tikzstyle{STATE}      =  [ fill=black!40!white, ]
\tikzstyle{INPUT}      =  [ fill=black!20!white, ]
\tikzstyle{UNCOD1}     =  [ pattern=north west lines ]
\tikzstyle{UNCOD2}     =  [ opacity=.4,fill=yellow!50!orange ]
\tikzstyle{UNCOD3}     =  [ opacity=.4,fill=yellow!20!orange ]
\tikzstyle{UNCOD1L}    =  [ draw=yellow!80!orange, ]
\tikzstyle{UNCOD2L}    =  [ draw=yellow!50!orange ]
\tikzstyle{UNCOD3L}    =  [ draw=yellow!20!orange ]
\tikzstyle{EXTEND}     =  [ pattern=north east lines, ]
\tikzstyle{TRRSSE}     =  [ pattern=north west lines, ]
\tikzstyle{XSB}        =  [ thick, |-|, shorten <=3pt, shorten >=3pt ]

\tikzset{orientation/.is choice,
    orientation/lr/.style={anchor=west,right=1},
    orientation/lr2/.style={anchor=west,right=2},
    orientation/lrd/.style={anchor=west,below=1},
    orientation/lrd2/.style={anchor=west,below=2},
    orientation/rl/.style={anchor=east,left=1},
    orientation/rl2/.style={anchor=east,left=2},
    orientation/ud/.style={anchor=north,below=1},
    orientation/du/.style={anchor=south,above=1},
    orientation/rld/.style={anchor=east,below=1},
    orientation/rld2/.style={anchor=east,below=2},
}
\tikzstyle{scare} = [
]
\tikzstyle{syslinear} = [
 drop shadow={shadow xshift=.6mm,
              shadow yshift=-.6mm},
 fill=white,
 anchor=west,
 rectangle,
 draw=black,
 minimum height=5mm,
 minimum width=5mm,
 inner xsep=0.5em
]

\tikzstyle{sysnonlinear} = [
 drop shadow={shadow xshift=.8mm,
              shadow yshift=-.8mm},
 fill=white,
 double,
 anchor=west,
 rectangle,
 draw=black,
 minimum height=5mm,
 minimum width=5mm,
 inner xsep=0.5em
]

\tikzstyle{syssource} = [
 anchor=west,
 ellipse,
 draw=black,
 minimum height=1.5ex,
 minimum width=1.5em,
 inner xsep=0.5em
]

\tikzstyle{syssink} = [
 anchor=west,
 ellipse,
 draw=black,
 minimum height=1.5ex,
 minimum width=1.5em,
 inner xsep=0.5em
]

\tikzstyle{syssplit} = [
 fill=black,
 draw=black,
]

\tikzstyle{sysadd} = [
 draw,circle,inner sep=-1pt,
]

\tikzstyle{sysmul} = [
 draw,circle,inner sep=-1pt,
]

\definecolor{MyHSBGreen}{hsb}{0.34065,1,0.91}
\pgfplotscreateplotcyclelist{colors4}{%
 {black!100!white},
 {black!75!white},
 {black!50!white},
 {black!25!white},
}
\pgfplotscreateplotcyclelist{colors5}{%
 {black!100!white},
 {black!75!white},
 {black!50!white},
 {black!25!white},
 {densely dashed, black!100!white},
 {densely dashed, black!75!white},
}
\pgfplotscreateplotcyclelist{colors10}{%
 {black!100!white},
 {black!75!white},
 {black!50!white},
 {black!25!white},
 {densely dashed, black!100!white},
 {densely dashed, black!75!white},
 {densely dashed, black!50!white},
 {densely dashed, black!25!white},
 {densely dotted, black!100!white},
 {densely dotted, black!75!white},
 {densely dotted, black!50!white},
}

\pgfplotscreateplotcyclelist{colorsBERMDDFSE}{%
                 every mark/.append style={fill=black},mark=o\\%
                 every mark/.append style={fill=black},mark=star\\%
                 every mark/.append style={fill=black},mark=triangle\\%
                 every mark/.append style={scale=.7,fill=black},mark=square\\%
                 every mark/.append style={fill=black},mark=diamond\\%
  densely dashed,every mark/.append style={solid},mark=pentagon*\\%
  densely dashed,every mark/.append style={solid},mark=*\\%
  densely dashed,every mark/.append style={solid},mark=triangle*\\%
  densely dashed,every mark/.append style={scale=.7,solid},mark=square*\\%
  densely dashed,every mark/.append style={solid},mark=diamond*\\%
  densely dashed,every mark/.append style={solid},mark=star*\\%
}
\pgfplotscreateplotcyclelist{colors1Empty4Full}{%
                 every mark/.append style={fill=black},mark=o\\%
  densely dashed,every mark/.append style={solid},mark=*\\%
  densely dashed,every mark/.append style={solid},mark=triangle*\\%
  densely dashed,every mark/.append style={scale=.7,solid},mark=square*\\%
  densely dashed,every mark/.append style={solid},mark=diamond*\\%
}

\pgfplotscreateplotcyclelist{colors4Empty4Full}{%
                 every mark/.append style={fill=black},mark=o\\%
                 every mark/.append style={fill=black},mark=triangle\\%
                 every mark/.append style={scale=.7,fill=black},mark=square\\%
                 every mark/.append style={fill=black},mark=diamond\\%
  densely dashed,every mark/.append style={solid},mark=*\\%
  densely dashed,every mark/.append style={solid},mark=triangle*\\%
  densely dashed,every mark/.append style={scale=.7,solid},mark=square*\\%
  densely dashed,every mark/.append style={solid},mark=diamond*\\%
}
\pgfplotscreateplotcyclelist{colors6Empty4Full}{%
                 every mark/.append style={fill=black},mark=o\\%
                 every mark/.append style={fill=black},mark=star\\%
                 every mark/.append style={fill=black},mark=triangle\\%
                 every mark/.append style={scale=.7,fill=black},mark=square\\%
                 every mark/.append style={fill=black},mark=diamond\\%
                 every mark/.append style={fill=black},mark=pentagon\\%
  densely dashed,every mark/.append style={solid},mark=*\\%
  densely dashed,every mark/.append style={solid},mark=triangle*\\%
  densely dashed,every mark/.append style={scale=.7,solid},mark=square*\\%
  densely dashed,every mark/.append style={solid},mark=diamond*\\%
}
\pgfplotscreateplotcyclelist{colors8Empty4Full}{%
                 every mark/.append style={fill=black},mark=o\\%
                 every mark/.append style={fill=black},mark=star\\%
                 every mark/.append style={fill=black},mark=triangle\\%
                 every mark/.append style={scale=.7,fill=black},mark=square\\%
                 every mark/.append style={fill=black},mark=diamond\\%
                 every mark/.append style={fill=black},mark=pentagon\\%
                 every mark/.append style={fill=black},mark=oplus\\%
                 every mark/.append style={fill=black},mark=otimes\\%
  densely dashed,every mark/.append style={solid},mark=*\\%
  densely dashed,every mark/.append style={solid},mark=triangle*\\%
  densely dashed,every mark/.append style={scale=.7,solid},mark=square*\\%
  densely dashed,every mark/.append style={solid},mark=diamond*\\%
}

\pgfplotscreateplotcyclelist{colors}{%
                 every mark/.append style={fill=black},mark=o\\%
                 every mark/.append style={fill=black},mark=star\\%
                 every mark/.append style={fill=black},mark=triangle\\%
                 every mark/.append style={scale=.7,fill=black},mark=square\\%
                 every mark/.append style={fill=black},mark=diamond\\%
                 every mark/.append style={fill=black},mark=pentagon\\%
                 every mark/.append style={fill=black},mark=oplus\\%
                 every mark/.append style={fill=black},mark=otimes\\%
                 every mark/.append style={fill=black},mark=asterisk\\%
                 every mark/.append style={fill=black},mark=+\\%
                 every mark/.append style={fill=black},mark=-\\%
                 every mark/.append style={fill=black},mark=|\\%
                 every mark/.append style={fill=black},mark=x\\%
                 every mark/.append style={fill=black},mark=*\\%
                 every mark/.append style={fill=black},mark=triangle*\\%
                 every mark/.append style={scale=.7,fill=black},mark=square*\\%
                 every mark/.append style={fill=black},mark=diamond*\\%
                 every mark/.append style={fill=black},mark=star*\\%
  densely dashed,every mark/.append style={solid,fill=gray},mark=o\\%
  densely dashed,every mark/.append style={solid,fill=gray},mark=triangle\\%
  densely dashed,every mark/.append style={scale=.7,solid,fill=gray},mark=square\\%
  densely dashed,every mark/.append style={solid,fill=gray},mark=diamond\\%
  densely dashed,every mark/.append style={solid,fill=gray},mark=star\\%
  densely dashed,every mark/.append style={solid,fill=gray},mark=pentagon*\\%
  densely dashed,every mark/.append style={solid,fill=gray},mark=*\\%
  densely dashed,every mark/.append style={solid,fill=gray},mark=triangle*\\%
  densely dashed,every mark/.append style={scale=.7,solid,fill=gray},mark=square*\\%
  densely dashed,every mark/.append style={solid,fill=gray},mark=diamond*\\%
  densely dashed,every mark/.append style={solid,fill=gray},mark=star*\\%
  mark=text,text mark=a\\%
  mark=text,text mark=b\\%
  mark=text,text mark=c\\%
  mark=text,text mark=d\\%
  mark=text,text mark=e\\%
  mark=text,text mark=f\\%
  mark=text,text mark=g\\%
}

\pgfdeclarelayer{background layer}
\pgfdeclarelayer{foreground layer}
\pgfsetlayers{background layer,main,foreground layer}

\newcommand{\ie}{\emph{i.e.}}
\newcommand{\eg}{\emph{e.g.}}
\newcommand{\cf}{\emph{cf.}}

\newcommand{\lvec}[1]{\ensuremath{\mathrm{\mathbf{#1}}}}  
\newcommand{\gvec}[1]{\ensuremath{\boldsymbol{#1}}}       

\providecommand{\de}[1]{\ensuremath{\mathop{\mathrm{d}}}}

\providecommand{\e}{\ensuremath{\mathrm{e}}}

\allowdisplaybreaks 

\IEEEoverridecommandlockouts
\title{Low Complexity Decoding for Punctured Trellis-Coded Modulation Over
       Intersymbol Interference Channels}
\author{
 \IEEEauthorblockN{Fabian~Schuh and
                   Johannes~B.~Huber}\\
 \IEEEauthorblockA{Institute for Information Transmission,
                   Friedrich-Alexander-Universit\"at Erlangen-N\"urnberg, Germany\\ 
                   mail: \texttt{\{schuh,\,huber\}@LNT.de}}%
}
\tikzstyle{EMTY}       =  [ fill=white, ]
\tikzstyle{STATE}      =  [ fill=black!40!white, ]
\tikzstyle{INPUT}      =  [ fill=black!20!white, ]
\tikzstyle{EXTEND}     =  [ pattern=north east lines, ]
\tikzstyle{TRRSSE}     =  [ pattern=north west lines, ]
\tikzstyle{XSB}        =  [ thick, |-|, shorten <=3pt, shorten >=3pt ]

\begin{document}
\sloppy
\maketitle
\begin{abstract}

Classical trellis-coded modulation (TCM) as introduced by Ungerboeck in
1976/1983 uses a signal constellation of twice the cardinality compared to an
uncoded transmission with one bit of redundancy per PAM symbol, \ie,
application of codes with rates $\frac{n-1}{n}$ when $2^{n}$ denotes the
cardinality of the signal constellation.
The original approach therefore only comprises integer transmission rates, \ie,
$R=\left\{ 2,\,3,\,4\,\ldots \right\}$,
additionally, when transmitting over an intersymbol interference (ISI) channel
an optimum decoding scheme would perform equalization and decoding of the
channel code jointly.

In this paper, we allow rate adjustment for TCM by means of puncturing the
convolutional code (CC) on which a TCM scheme is based on. In this case a
nontrivial mapping of the output symbols of the CC to signal points results in
a time-variant trellis.
We propose an efficient technique to integrate an ISI-channel into this trellis
and show that the computational complexity can be significantly reduced by
means of a reduced state sequence estimation (RSSE) algorithm for time-variant
trellises.
\end{abstract}

\begin{IEEEkeywords}
trellis-coded modulation (TCM);
punctured convolutional codes;
Viterbi-Algorithm (VA);
ISI-channel;
\end{IEEEkeywords}
\IEEEpeerreviewmaketitle
\vspace*{-3ex}
\section{Introduction                                         } Ungerboeck's trellis-coded modulation (TCM)~\cite{UngerboeckTCM87} is a
bandwidth efficient digital transmission scheme when very low overall latency
is desired. Low latency is ensured by the use of convolutional codes instead of
block codes (\cf~\cite{LIT_tr_com_2009_hehn}) and the dispense with
interleaving (as opposed to conventional bit-interleaved coded
modulation~\cite{141453}).

Ungerboeck showed that a significant increase in the Asymptotic Coding Gain
(ACG) can be achieved when considering channel coding and modulation jointly.
By expanding the constellation from $2^{n-1}$ to $2^{n}$ signal points and
employing a rate-$\frac{n-1}{n}$ convolutional encoder one can improve the
robustness of the transmission against noise by up to $6\,$dB without any
further costs besides some computational effort. However, TCM is strictly
limited to integer transmission rates.

Our approach applies \emph{punctured} TCM (P-TCM) with an arbitrary rate. We
extend P-TCM to intersymbol interference (ISI)-channel scenarios. In this case,
ML-decoding can be performed by efficiently incorporating the ISI-channel into
the trellis.

We show that reduced-state sequence estimation (RSSE) can be applied in order
to reduce computational complexity. We were able to show
in~\cite{Schu1301:Reduced,Schu1301:Matched} that for minimum phase channels,
the number of states to decode must \emph{not} be significantly higher than the
number of states in the channel encoder. In this paper we will describe the
application of RSSE to P-TCM and discuss the partitioning of the time-variant
trellis into hyperstates.

This paper is structured as follows: In Sec.~\ref{sec:systemmodell} we first
introduce notation and present the system model. Sec.~\ref{sec:ptcm} briefly
recapitulates a presentation technique that enables the implementation of
punctured encoding. The application of RSSE for P-TCM is given in
Sec.~\ref{sec:RSSE}. Final results of numerical simulation and conclusions are
given in Sec.~\ref{sec:results} and Sec.~\ref{sec:conclusion}, respectively.
 
\section{System Model                                         } \label{sec:systemmodell}

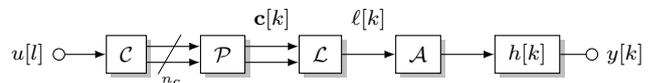
\begin{figure}[ht]\vspace*{-2ex}
 \begin{center}
  \begin{tikzpicture}[>=latex,x=1em,y=4ex,font=\footnotesize,inner sep=0.3em,
                      node distance=10mm and 4mm]
   \node at (0,0) (u) {$u[l]$};
   \node[syslinear,right=7mm,at=(u.east)] (Encoding) {$\mathcal{C}$};
   \node[syslinear,right=7mm,at=(Encoding.east)] (Puncturing) {$\mathcal{P}$};
   \node[syslinear,right=7mm,at=(Puncturing.east)] (Labeling) {$\mathcal{L}$};
   \node[syslinear,right=7mm,at=(Labeling.east)] (Mapper) {$\mathcal{A}$};
   \node[syslinear,right=7mm,at=(Mapper.east)] (ISI) {$h[k]$};
   \draw[->] ($(Encoding.east)+(0,3pt)$) -- ($(Puncturing.west)+(0,3pt)$);
   \draw[->] ($(Encoding.east)-(0,3pt)$) -- ($(Puncturing.west)-(0,3pt)$);
   \draw[->] ($(Puncturing.east)+(0,3pt)$) -- node[midway,above=0.7ex] {$\mathbf{c}[k]$} ($(Labeling.west)+(0,3pt)$);
   \draw[->] ($(Puncturing.east)-(0,3pt)$) -- ($(Labeling.west)-(0,3pt)$);
   \draw[->] (Labeling.east) -- node[pos=.5,above=1.3ex] {$\ell[k]$} (Mapper);
   \draw[->] (Mapper.east) -- (ISI);
   \draw[o->] (u) -- (Encoding);
   \draw[-o] (ISI.east) -- ++(5mm,0) node[right] {$y[k]$};
   \draw[very thin] (Encoding.east) ++(+3mm,0) +(240:3mm) node[below right,font=\tiny,inner sep=1pt] {$n_\mathrm{c}$} -- ++(60:3mm);
  \end{tikzpicture}
 \end{center}\vspace*{-3ex}
 \caption{Concatenation of a rate-$\frac12$ convolutional encoder $\mathcal{C}$
          and puncturing $\mathcal{P}$ with labeling and modulation
          ($n_\mathrm{u}=0$, $n_\mathrm{c}=2$).}
 \label{fig:sysmodel}
 \vspace*{-2ex}
\end{figure}

This paper deals with convolutionally encoded pulse-amplitude modulated
(PAM) transmission as depicted in Fig.~\ref{fig:sysmodel}. (Here, the term
PAM is used for complex-valued signal constellations $\mathcal{A}$ as well
including amplitude-shift keying (ASK), phase-shift keying (PSK) or
quadrature-amplitude modulation (QAM).)
A binary data sequence $\langle u\rangle$ is encoded using a
rate-$\frac{n_\mathrm{c}-1}{n_\mathrm{c}}$ binary convolutional encoder
$\mathcal{C}$ with generator polynomials $g_{ij}(D),\; 1 \leq i \leq
n_\mathrm{c}; 1 \leq j \leq n_\mathrm{c}-1$, with delay operator $D$,
$n_\mathrm{c}-1$ parallel binary-input symbols and $n_\mathrm{c}$ parallel
output symbols at each time instant.

At each output of the encoder, the symbols traverse through a puncturing system
with puncturing matrix $\mathbf{P} = [P_{ij}],\,P_{ij}\in\left\{ 0,\,1 \right\};\;1<i\leq
n_\mathrm{c};\;1<j<\Omega$ and period $\Omega$. For each
($n_\mathrm{c}$)-tuple of encoder output symbols the puncturing scheme
cyclically advances by one step. Where $P_{ij}$ is zero, the current symbol at
the output is discarded, accordingly.

The punctured encoded output symbols $\mathbf{c}[k]$ are labeled to $\ell[k]$ before
being mapped to the $M=2^{n_\mathrm{u}+n_\mathrm{c}}=2^n$-ary signal
constellation $\mathcal{A}$.

The modulated (possibly complex-valued) transmit signal traverses through a
memory-$L$ discrete-time ISI-channel with $L+1$ channel coefficients $h[k]$
with $k$ denoting the discrete time index.

The task of the receiver is to estimate for the information bits given the
transmit signal $y[k]$ plus additive noise. Here, we focus on perfect
channel knowledge at the receiver-side.
 
\section{Punctured Trellis-Coded Modulation                   } \label{sec:ptcm}
In the following, we will briefly recapitulate punctured convolutional trellis
coded transmission over ISI-channel scenarios as introduced
in~\cite{Schu1301:Matched}.

In contrast to classical TCM, our approach using \emph{punctured} convolutional
codes results in nontrivial mapping of coded bits to modulation symbols. As a
consequence, the trellis is time-variant as already described
in~\cite{Schu1301:Matched,502012,PTCMUnderReviewICC14}. 

\begin{figure}[ht]\vspace*{-2ex}
 \begin{center}
  \begin{tikzpicture}[>=latex,x=6em,y=3ex,font=\footnotesize,inner sep=0.3em,
                      node distance=10mm and 4mm]
   \coordinate (u0) at (0,0);
   \coordinate (u1) at (1,0);
   \coordinate (u2) at (2,0);
   \coordinate (u3) at (3,0);
   \draw (u0) circle (1pt) node[above] {$u[l]$};
   \draw (u1) circle (1pt) node[above] {$u[l+1]$};
   \draw (u2) circle (1pt) node[above] {$u[l+2]$};
   \draw (u3) circle (1pt) node[above] {$u[l+3]$};
   \draw (-0.25,-1) circle (1pt) node[coordinate] (c0) {};
   \draw (+0.25,-1) circle (1pt) node[coordinate] (c1) {};
   \draw (+0.75,-1) circle (1pt) node[coordinate] (c2) {};
   \draw (+1.25,-1) circle (1pt) node[coordinate] (c3) {};
   \draw (+1.75,-1) circle (1pt) node[coordinate] (c4) {};
   \draw (+2.25,-1) circle (1pt) node[coordinate] (c5) {};
   \draw (+2.75,-1) circle (1pt) node[coordinate] (c6) {};
   \draw (+3.25,-1) circle (1pt) node[coordinate] (c7) {};
   \begin{scope}[shorten <=1pt,shorten >=1pt]
    \draw[->] (u0) -- node[midway,above,sloped] {$g_1$} (c0);
    \draw[->] (u0) -- node[midway,above,sloped] {$g_2$} (c1);
    \draw[->] (u1) -- node[midway,above,sloped] {$g_1$} (c2);
    \draw[->] (u1) -- node[midway,above,sloped] {$g_2$} (c3);
    \draw[->] (u2) -- node[midway,above,sloped] {$g_1$} (c4);
    \draw[->] (u2) -- node[midway,above,sloped] {$g_2$} (c5);
    \draw[->] (u3) -- node[midway,above,sloped] {$g_1$} (c6);
    \draw[->] (u3) -- node[midway,above,sloped] {$g_2$} (c7);
    \draw (-0.25,-2) node[draw] (s0) {MSB};
    \draw (+0.25,-2) node[draw] (s1) {LSB};
    \draw (+0.75,-2) node       (s2) {\mbox{ }};
    \draw (+1.25,-2) node[draw] (s3) {MSB};
    \draw (+1.75,-2) node[draw] (s4) {LSB};
    \draw (+2.25,-2) node[draw] (s5) {MSB};
    \draw (+2.75,-2) node       (s6) {\mbox{ }};
    \draw (+3.25,-2) node[draw] (s7) {LSB};
   \end{scope}
   \node at (c2) {\Large\textbf{$\times$}};
   \node at (c6) {\Large\textbf{$\times$}};
   \begin{scope}[shorten <=1pt,shorten >=1pt]
    \foreach \x in {0,1,3,4,5,7} {
     \draw[->] (c\x) -- (s\x);
    }
   \end{scope}
   \draw ($(s0.south west)-(1pt,1pt)$) -| ($(s1.north east)+(1pt,1pt)$) -| ($(s0.south west)-(1pt,1pt)$);
   \draw ($(s3.south west)-(1pt,1pt)$) -| ($(s4.north east)+(1pt,1pt)$) -| ($(s3.south west)-(1pt,1pt)$);
   \draw ($(s5.south west)-(1pt,1pt)$) -| ($(s7.north east)+(1pt,1pt)$) -| ($(s5.south west)-(1pt,1pt)$);
   \begin{scope}[decoration={brace,amplitude=.5em},decorate]
    \draw[decorate] ($(s1.south east)-(0,2pt)$) -- node[yshift=-1ex,midway,below] (l0) {$\ell[k]  $} ($(s0.south west)-(0,2pt)$);
    \draw[decorate] ($(s4.south east)-(0,2pt)$) -- node[yshift=-1ex,midway,below] (l1) {$\ell[k+1]$} ($(s3.south west)-(0,2pt)$);
    \draw[decorate] ($(s7.south east)-(0,2pt)$) -- node[yshift=-1ex,midway,below] (l2) {$\ell[k+2]$} ($(s5.south west)-(0,2pt)$);
    \draw[->] (l0) -- ++(0,-3ex) node[below] {$a[k]$};
    \draw[->] (l1) -- ++(0,-3ex) node[below] {$a[k+1]$};
    \draw[->] (l2) -- ++(0,-3ex) node[below] {$a[k+2]$};
   \end{scope}
   \draw[densely dashed,very thick,gray!80!white,shorten <=1ex] (l1.north) -- ++(0,2cm) node[coordinate] (repScheme) {};
   \path (s0.west) -- +(0cm,1cm) coordinate (foobar1);
   \path (repScheme)-- +(-1cm,0cm) coordinate (foobar2);
   \path (intersection cs: first line={(s0.west) -- (foobar1)}, second line={(repScheme)-- (foobar2)}) coordinate (foobar3);
   \draw[<->,shorten <=1em,shorten >=1em] (foobar3) -- node[midway,fill=white] {$P$} (repScheme);
   \path (s7.east) -- +(0cm,1cm) coordinate (foobar1);
   \path (repScheme)-- +(1cm,0cm) coordinate (foobar2);
   \path (intersection cs: first line={(s7.east) -- (foobar1)}, second line={(repScheme)-- (foobar2)}) coordinate (foobar3);
   \draw[<->,shorten <=1em,shorten >=1em] (foobar3) -- node[midway,fill=white] {$P$} (repScheme);
  \end{tikzpicture}
 \end{center}\vspace*{-3ex}
 \caption{Encoding process for a rate-$\frac23$ punctured convolutional
          code and natural labeling. Overall transmission rate
          $R=\frac43$.}
 \label{fig:convEncodingPuncturedProgress}
 \vspace*{-2ex} 
\end{figure}
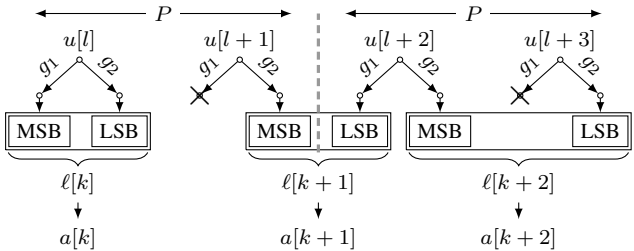

In order to briefly recapitulate decoding concept for punctured trellis coded
modulation, we focus on $4$-ary ASK-modulation, a memory-$2$ convolutional
code, and a short puncturing scheme namely $\mathbf{P} =
\left[\,(1\,1)^\top\,(0\,1)^\top\,\right]$.
Note that, whenever the number of erased bits in one period of the puncturing
scheme is not dividable by $\log_2(M)$, the puncturing scheme has to be
repeated until this condition is fulfilled. This restriction ensures that
entire modulation symbols can be constructed by the finite state machine (FSM).
In our example, the puncturing period (\eg,
$\left[\,(1\,1)^\top\,(0\,1)^\top\,\right]$) has to be applied twice. As can be
seen from the encoding process in Fig.~\ref{fig:convEncodingPuncturedProgress},
the third and the seventh encoded symbol are punctured and do not contribute to
the labeling and modulation process. Thus, the second symbol, \ie, $a[k+1]$,
contains information about $u[l+1]$ and $u[l+2]$ and the third symbol, \ie,
$a[k+2]$, contains information about $u[l+2]$ and $u[l+3]$.

\begin{figure}[ht]\vspace*{-2ex}
 \begin{center}
  \begin{tikzpicture}[x=25mm,y=4mm]
   \draw[fill] ( 1,7) circle ( 1pt) -- ( 2,7) circle ( 1pt);
   \draw[fill] ( 1,6) circle ( 1pt) -- ( 2,7) circle ( 1pt);
   \draw[fill] ( 1,5) circle ( 1pt) -- ( 2,6) circle ( 1pt);
   \draw[fill] ( 1,4) circle ( 1pt) -- ( 2,6) circle ( 1pt);
   \draw[fill] ( 1,7) circle ( 1pt) -- ( 2,5) circle ( 1pt);
   \draw[fill] ( 1,6) circle ( 1pt) -- ( 2,5) circle ( 1pt);
   \draw[fill] ( 1,5) circle ( 1pt) -- ( 2,4) circle ( 1pt);
   \draw[fill] ( 1,4) circle ( 1pt) -- ( 2,4) circle ( 1pt);
   \draw[fill] ( 2,7) circle ( 1pt) -- ( 3,7) circle ( 1pt);
   \draw[fill] ( 2,6) circle ( 1pt) -- ( 3,7) circle ( 1pt);
   \draw[fill] ( 2,5) circle ( 1pt) -- ( 3,6) circle ( 1pt);
   \draw[fill] ( 2,4) circle ( 1pt) -- ( 3,6) circle ( 1pt);
   \draw[fill] ( 2,3) circle ( 1pt) -- ( 3,5) circle ( 1pt);
   \draw[fill] ( 2,2) circle ( 1pt) -- ( 3,5) circle ( 1pt);
   \draw[fill] ( 2,1) circle ( 1pt) -- ( 3,4) circle ( 1pt);
   \draw[fill] ( 2,0) circle ( 1pt) -- ( 3,4) circle ( 1pt);
   \draw[fill] ( 2,7) circle ( 1pt) -- ( 3,3) circle ( 1pt);
   \draw[fill] ( 2,6) circle ( 1pt) -- ( 3,3) circle ( 1pt);
   \draw[fill] ( 2,5) circle ( 1pt) -- ( 3,2) circle ( 1pt);
   \draw[fill] ( 2,4) circle ( 1pt) -- ( 3,2) circle ( 1pt);
   \draw[fill] ( 2,3) circle ( 1pt) -- ( 3,1) circle ( 1pt);
   \draw[fill] ( 2,2) circle ( 1pt) -- ( 3,1) circle ( 1pt);
   \draw[fill] ( 2,1) circle ( 1pt) -- ( 3,0) circle ( 1pt);
   \draw[fill] ( 2,0) circle ( 1pt) -- ( 3,0) circle ( 1pt);
   \draw[fill] ( 3,7) circle ( 1pt) -- ( 4,7) circle ( 1pt);
   \draw[fill] ( 3,6) circle ( 1pt) -- ( 4,7) circle ( 1pt);
   \draw[fill] ( 3,5) circle ( 1pt) -- ( 4,7) circle ( 1pt);
   \draw[fill] ( 3,4) circle ( 1pt) -- ( 4,7) circle ( 1pt);
   \draw[fill] ( 3,3) circle ( 1pt) -- ( 4,6) circle ( 1pt);
   \draw[fill] ( 3,2) circle ( 1pt) -- ( 4,6) circle ( 1pt);
   \draw[fill] ( 3,1) circle ( 1pt) -- ( 4,6) circle ( 1pt);
   \draw[fill] ( 3,0) circle ( 1pt) -- ( 4,6) circle ( 1pt);
   \draw[fill] ( 3,7) circle ( 1pt) -- ( 4,5) circle ( 1pt);
   \draw[fill] ( 3,6) circle ( 1pt) -- ( 4,5) circle ( 1pt);
   \draw[fill] ( 3,5) circle ( 1pt) -- ( 4,5) circle ( 1pt);
   \draw[fill] ( 3,4) circle ( 1pt) -- ( 4,5) circle ( 1pt);
   \draw[fill] ( 3,3) circle ( 1pt) -- ( 4,4) circle ( 1pt);
   \draw[fill] ( 3,2) circle ( 1pt) -- ( 4,4) circle ( 1pt);
   \draw[fill] ( 3,1) circle ( 1pt) -- ( 4,4) circle ( 1pt);
   \draw[fill] ( 3,0) circle ( 1pt) -- ( 4,4) circle ( 1pt);
   \draw[shorten <=3pt,shorten >=3pt,|-|] (1,8) -- node[midway,circle,inner sep=2pt,fill=white,draw] {$\Gamma_0$} (2,8);
   \draw[shorten <=3pt,shorten >=3pt,|-|] (2,8) -- node[midway,circle,inner sep=2pt,fill=white,draw] {$\Gamma_1$} (3,8);
   \draw[shorten <=3pt,shorten >=3pt,|-|] (3,8) -- node[midway,circle,inner sep=2pt,fill=white,draw] {$\Gamma_2$} (4,8);
  \end{tikzpicture}\vspace*{-3ex}
 \end{center}
 \caption{Time-variant trellis for a punctured rate-$\nicefrac23$
          convolutional code. In the first to VA steps, two transitions arrive
          at each state, \ie, one bit can be estimate, whereas the third step
          allows an estimation for two bits.}
 \label{fig:nonLinearTrellis}
 \vspace*{-2ex}
\end{figure}
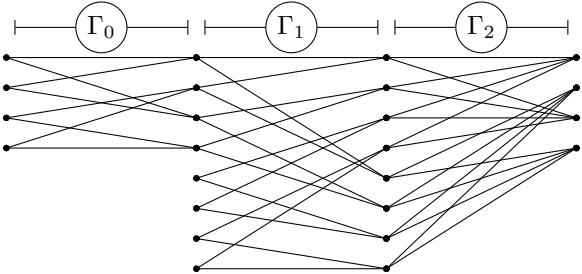

When decoding the second symbol (\eg, $a[k+1]$), a decision can be made for
$u[l+1]$ but not for $u[l+2]$, as a portion of information will be received in
the consecutive symbol. Thus, the trellis has to be expanded in order to use
the symbols $a[k+1]$ and $a[k+2]$ when decoding $u[l+2]$. As sample trellis is
given in Fig.~\ref{fig:nonLinearTrellis}.

To algorithmically handle the time-variant mapping we introduced a set of
so-called generator offsets $\mathcal{T}_i$ which describe, depending on the
puncturing scheme, modulation size, and time instant, the relations between
generator polynomials, input value, FSM state, and mapping to MSB or LSB,
respectively. For each new generator offset $\mathcal{T}_i$ a new trellis
segment arises, \eg, the number of generator offsets equals the number of
trellis segments in one trellis period.

\begin{figure}
 \begin{center}
  \begin{tikzpicture}
   \tikzset{row 1 column 8/.style={nodes={STATE}}}
   \tikzset{row 1 column 7/.style={nodes={STATE}}}
   \tikzset{row 1 column 6/.style={nodes={STATE}}}
   \tikzset{row 1 column 5/.style={nodes={STATE}}}
   \tikzset{row 1 column 4/.style={nodes={INPUT}}}
   \tikzset{row 2 column 6/.style={nodes={STATE}}}
   \tikzset{row 2 column 5/.style={nodes={STATE}}}
   \tikzset{row 2 column 4/.style={nodes={STATE}}}
   \tikzset{row 2 column 3/.style={nodes={STATE}}}
   \tikzset{row 2 column 2/.style={nodes={INPUT}}}
   \tikzset{row 3 column 5/.style={nodes={STATE}}}
   \tikzset{row 3 column 4/.style={nodes={STATE}}}
   \tikzset{row 3 column 3/.style={nodes={STATE}}}
   \tikzset{row 3 column 2/.style={nodes={STATE}}}
   \tikzset{row 3 column 1/.style={nodes={INPUT}}}
   \matrix (FIFO) [matrix of nodes,
                   nodes in empty cells,
                   nodes={draw,
                          ultra thin,
                          anchor=south,
                          rectangle,
                          text width=1.7em,
                          minimum height=6ex,
                         },
                   ] {
     &&&&&&&\\
     &&&&&&&\\
     &&&&&&&\\
    };
    \draw[XSB] ($(FIFO-1-4.west)+(0,2ex)$)            -- node[midway,above,font=\tiny,text=white,inner sep=.1pt] {MSB} ($(FIFO-1-6.east)+(0,2ex)$);
    \draw[XSB] ($(FIFO-1-4.west)+(0,1ex)$)            -- node[midway,above,font=\tiny,text=white,inner sep=.1pt] {LSB} ($(FIFO-1-6.east)+(0,1ex)$);
    \draw[XSB,opacity=.5] ($(FIFO-1-6.west)-(0,1ex)$) -- ($(FIFO-1-8.east)-(0,1ex)$);
    \draw[XSB,opacity=.5] ($(FIFO-1-5.west)-(0,2ex)$) -- ($(FIFO-1-7.east)-(0,2ex)$);
    \draw[XSB] ($(FIFO-2-3.west)+(0,2ex)$)            -- ($(FIFO-2-5.east)+(0,2ex)$);
    \draw[XSB] ($(FIFO-2-2.west)+(0,1ex)$)            -- ($(FIFO-2-4.east)+(0,1ex)$);
    \draw[XSB,opacity=.5] ($(FIFO-2-4.west)-(0,1ex)$) -- ($(FIFO-2-6.east)-(0,1ex)$);
    \draw[XSB,opacity=.5] ($(FIFO-2-4.west)-(0,2ex)$) -- ($(FIFO-2-6.east)-(0,2ex)$);
    \draw[XSB] ($(FIFO-3-2.west)+(0,2ex)$)            -- ($(FIFO-3-4.east)+(0,2ex)$);
    \draw[XSB] ($(FIFO-3-1.west)+(0,1ex)$)            -- ($(FIFO-3-3.east)+(0,1ex)$);
    \draw[XSB,opacity=.5] ($(FIFO-3-3.west)-(0,1ex)$) -- ($(FIFO-3-5.east)-(0,1ex)$);
    \draw[XSB,opacity=.5] ($(FIFO-3-2.west)-(0,2ex)$) -- ($(FIFO-3-4.east)-(0,2ex)$);
    \node[font=\footnotesize,right=1pt,ellipse,draw,inner sep=0pt]            at ($(FIFO-1-8.east)+(0,1.5ex)$) {$h[0]$};
    \node[font=\footnotesize,right=1pt,ellipse,draw,inner sep=0pt,opacity=.5] at ($(FIFO-1-8.east)+(0,-1.5ex)$) {$h[1]$};
    \node[font=\footnotesize,right=1pt,ellipse,draw,inner sep=0pt]            at ($(FIFO-2-6.east)+(0,1.5ex)$) {$h[0]$};
    \node[font=\footnotesize,right=1pt,ellipse,draw,inner sep=0pt,opacity=.5] at ($(FIFO-2-6.east)+(0,-1.5ex)$) {$h[1]$};
    \node[font=\footnotesize,right=1pt,ellipse,draw,inner sep=0pt]            at ($(FIFO-3-5.east)+(0,1.5ex)$) {$h[0]$};
    \node[font=\footnotesize,right=1pt,ellipse,draw,inner sep=0pt,opacity=.5] at ($(FIFO-3-5.east)+(0,-1.5ex)$) {$h[1]$};
    \node[left] at (FIFO-1-1.west) {\parbox{3em}{$\mathcal{T}_0$\\$\hspace*{1em}\mathcal{T}_2$}};
    \node[left] at (FIFO-2-1.west) {\parbox{3em}{$\mathcal{T}_1$\\$\hspace*{1em}\mathcal{T}_0$}};
    \node[left] at (FIFO-3-1.west) {\parbox{3em}{$\mathcal{T}_2$\\$\hspace*{1em}\mathcal{T}_1$}};
    \node[above,font=\footnotesize] at (FIFO-1-1.north) {$k+3$};
    \node[above,font=\footnotesize] at (FIFO-1-2.north) {$k+2$};
    \node[above,font=\footnotesize] at (FIFO-1-3.north) {$k+1$};
    \node[above,font=\footnotesize] at (FIFO-1-4.north) {$k$};
    \node[above,font=\footnotesize] at (FIFO-1-5.north) {$k-1$};
    \node[above,font=\footnotesize] at (FIFO-1-6.north) {$k-2$};
    \node[above,font=\footnotesize] at (FIFO-1-7.north) {$k-3$};
    \node[above,font=\footnotesize] at (FIFO-1-8.north) {$k-4$};
  \end{tikzpicture}
 \end{center}\vspace*{-3ex}
 \caption{State transitions of the transmitter FSM with $R=\nicefrac43$ and the
          relations between generator polynomials, FSM-state/input and channel
          state for a memory-$1$ ISI-channel.}
 \label{fig:MatchedPuncturedConvISI}
 \vspace*{-2ex}
\end{figure}
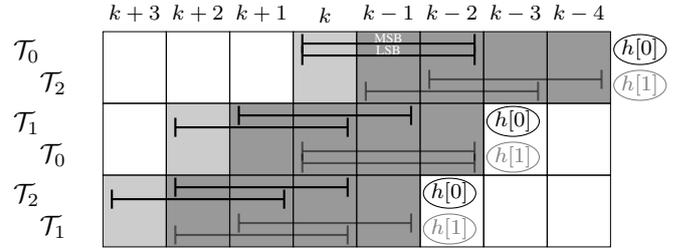

When transmitting over an ISI-channel, several symbols are stored in the memory
of the ISI-channel independently from the encoding and puncturing process.
Thus, multiple generator offsets $\mathcal{T}_i$ have to be considered
simultaneously. This can be seen from Fig.~\ref{fig:MatchedPuncturedConvISI}
for a memory-$1$ channel. There, the resulting sequence of used generator
offsets is depicted. This scheme can easily be extended to arbitrary lengths
of the ISI-channel. A detailed description as well as an algorithm to
construct such trellises will be given in a separate full-length paper.

\section{Reduced-State Sequence Estimation                    } \label{sec:RSSE}
Reduced-state sequence estimation (RSSE)~\cite{Eyuboglu1988} is proposed to
reduce the number of states at the cost of small loss in Euclidean distance. In
order to introduce RSSE for P-TCM, we first briefly recapitulate \emph{delayed
decision-feedback sequence estimation} (DFSE)~\cite{Duel-Hallen1989}.
 
\subsection{Delayed Decision Feedback Sequence Estimation     } When equalizing (uncoded) digital PAM signaling over a discrete-time
ISI-channel with $L+1$ tabs using DFSE (\ie, no decoding), the trellis is
constructed from the first $\tilde{L}\leq L$ tabs only. Thus, the number of
states is reduced from $M^L$ to $M^{\tilde{L}}$.

The remaining $L+1-\tilde{L}$ channel tabs are considered in a delayed
decision-feedback equalization (DFE) that is performed in each trellis state
using the \emph{delayed} path register of the corresponding state.

The main difference to full state equalization appears in the metric
computation for each time instant. From~(\ref{eq:dsfeMetric}) it becomes clear
that the state specific path register $p_\mathrm{reg}[k,\lvec{s}]$ is delayed
by $\tilde{L}$ and its elements are multiplied by the subsequent channel
coefficients $h_\mathrm{dfe}[h]$ which have not been considered in the trellis.
The branch metric $\lambda(\lvec{s},\lvec{u})$ (\eg, Euclidean distance of the
received symbol $y[k]$ to the hypotheses $h(\lvec{s},\lvec{u})$ for the states
$\lvec{s}$ and bits $\lvec{u}$) thus includes the correction factor $\delta$:
\begin{align}\label{eq:dsfeMetric}
 \delta                     & = \sum\limits_\kappa p_\mathrm{reg}[k-\tilde{L}+\kappa,\lvec{s}]\,\cdot\,h_\mathrm{dfse}[\kappa]\\
 \lambda(\lvec{s},\lvec{u}) & = \big| y[k] - h(\lvec{s},\lvec{u}) - \delta\big|^2\notag
\end{align}
 
\subsection{Reduced State Sequence Estimation                 } For our coded transmission over ISI-channel we consider RSSE instead. Here, $Z$
arbitrary MLSE states, each with $M=2^K$ possible branches to adjacent states,
are combined into $Z_\text{R} = \frac{Z}{2^J};\;J\in\mathbb{N}$
\emph{hyperstates}~\cite{Spinnler1995} each having $2^J$ substates and
$2^K\cdot2^J$ branches. A certain assignment of states to hyperstates is called
a \textit{partitioning}~\cite{Spinnler1995}.

Instead of having $2^K$ arriving branches at each of the $Z$ MLSE states we get
a set of $2^K\cdot2^J$ branches at each of the $Z_\text{R}$ hyperstates. The
total number of available branches remains $2^K\cdot Z$. However, when using
RSSE only $2^K$ branches are possible (\ie, \emph{enabled}) from each state, at
a given time instant. The availability of branches is determined by the path
registers, and thus form a decision-feedback.

The metric computation in this case can be implemented as depicted in
Algorithm~\ref{alg:MetricRSSE}. Note that with line~\ref{alg:MetricRSSEactive}
only $M$ branches are activated. Thus, the VA has to decide between $M$
survivor branches at each state giving an estimate for $\log_2(M)$ bits.

\begin{algorithm}
 \caption{Metric calculations for RSSE}
 \label{alg:MetricRSSE}
 \begin{algorithmic}[1]
  \State $\tilde{L} \gets \log_2(\mathrm{nr.~hyperstates})/\log_2(M)$
  \State $\ell \gets \log_2(\mathrm{nr.~substates})/\log_2(M)$
  \ForAll{$\lvec{s} \in \mathcal{S}$}
   \ForAll{$\lvec{u} \in \mathcal{A} \cdot K$}
    \ForAll{$\kappa =1 \to \ell$}
     \State $\zeta[\kappa] =
     p_\mathrm{reg}(\mathrm{end}-\tilde{L}+\kappa,\lvec{s})$ \Comment{active branches}\label{alg:MetricRSSEactive}
    \EndFor
    \State$\displaystyle\lambda(\lvec{s},\lvec{u}) \gets \big| y[k] - h(\gvec{\zeta},\lvec{u})\big|^2$\Comment{branch metric}
    \State$\lambda(\lvec{s},\lvec{u}) \gets \lambda(\lvec{s},\lvec{u}) + \Gamma(\lvec{s})$\Comment{acc. path metric}
   \EndFor
  \EndFor
 \end{algorithmic}
\end{algorithm}

For time-variant trellises some modifications to the underlying VA are
necessary, which are described in~\cite{Schu1301:Matched}.

The main difference to MLSE is, that we decide for a surviving path prematurely
resulting in a truncation of error events. A loss in Euclidean distance appears
if an error event with minimum Euclidean distance gets truncated. Therefore the
performance of RSSE strongly depends on the partitioning of the states into
hyperstates. Instead of exhaustively search for the optimum state
partitioning, which maximizes the intra-hyperstate
distance~\cite{Spinnler1995}, we exploit the minimum phase characteristics of
the ISI-channel which is, as described above, fully integrated into our trellis.

For a minimum phase channel impulse response the prior channel input symbols
are weighted less than more recent ones and, thus, affect the metric less.  The
elder the symbols, the further back it is stored in the vector presentation of
a particular trellis state. Hence, the intra-hyperstate distance is maximized
when states are combined with respect to elder positions in the state number.
This particular partitioning is equivalent to DFSE for ISI-channels (which is
the optimum partitioning for equalization of minimum phase
ISI-channels~\cite{Spinnler1995}) and will in the latter be called \emph{DFSE
partitioning}. As the minimum phase ISI-channel is the last element to affect
the transmitted symbols and is also fully integrated into the trellis, we can
apply the \emph{DFSE partitioning} to use RSSE for P-TCM over ISI-channels.
An implementation of this set partitioning for the $J$\textsuperscript{th}
level exploiting the minimum phase characteristics is shown in
Algorithm~\ref{alg:partminphase}. The columns in the resulting matrix $p(z,i)$
define the states that have to be grouped into hyperstates.

\begin{algorithm}
 \caption{DFSE Partitioning for RSSE}
 \label{alg:partminphase}
 \begin{algorithmic}[1]
  \State $i,z \gets 1$
  \For{$\ell = 1 \to Z$}
   \If{$\ell \operatorname{mod} J = 0$}
    \State $z\gets z+1$
   \EndIf
   \State $p(z,i) = \ell$
   \State $i\gets i+1$
  \EndFor
 \end{algorithmic}
\end{algorithm}

In the following we will focus on our state design and two possibilities to
apply \emph{DFSE partitioning} to time-variant trellises.

\subsection{State Design                                      } \label{sec:decoding}

\begin{figure}[ht]\vspace*{-2ex}
 \begin{center}
  \begin{tikzpicture}
   \tikzset{row 1 column 1/.style={nodes={EXTEND}}}
   \tikzset{row 1 column 2/.style={nodes={STATE}}}
   \tikzset{row 1 column 3/.style={nodes={STATE}}}
   \tikzset{row 1 column 4/.style={nodes={STATE}}}
   \tikzset{row 1 column 5/.style={nodes={STATE}}}
   \tikzset{row 1 column 6/.style={nodes={STATE}}}
   \matrix (FIFO) [matrix of nodes,
                   nodes in empty cells,
                   nodes={draw,
                          ultra thin,
                          anchor=south,
                          rectangle,
                          text width=1.5em,
                          minimum height=3ex,
                         },
                   ] {
     &&&&&&&&\\
    };
    \begin{scope}[|-|,shorten >=1pt,shorten <=1pt,inner sep=1pt]
     \draw[|-,densely dotted,shorten >=0pt] ($(FIFO-1-1.north west)+(0,1ex)$) -- ($(FIFO-1-1.north east)+(0,1ex)$);
     \draw ($(FIFO-1-2.north west)+(0,1ex)$) -- node[above,font=\footnotesize,midway] {MLSE states}   ($(FIFO-1-6.north east)+(0,1ex)$);
     \draw ($(FIFO-1-7.north west)+(0,1ex)$) -- node[above,font=\footnotesize,midway] {path register} ($(FIFO-1-9.north east)+(0,1ex)$);
    \end{scope}
    \draw[->] (FIFO-1-9.south) ++(0,-1.5ex) node[right] {$t$} -- ++(-1em,0);
    \begin{scope}[line width=.1pt]
     \draw (FIFO-1-1.west) -- node[midway,fill=white,inner sep=1pt,font=\tiny] {$0$} ++(160:1cm) circle ( 1pt);
     \draw (FIFO-1-1.west) -- node[midway,fill=white,inner sep=1pt,font=\tiny] {$1$} ++(200:1cm) circle ( 1pt);
     \draw[fill] (FIFO-1-1.west) circle ( 1pt);
    \end{scope}
  \end{tikzpicture}
  \begin{tikzpicture}
   \tikzset{row 1 column 1/.style={nodes={EXTEND}}}
   \tikzset{row 1 column 2/.style={nodes={STATE}}}
   \tikzset{row 1 column 3/.style={nodes={STATE}}}
   \tikzset{row 1 column 4/.style={nodes={STATE}}}
   \tikzset{row 1 column 5/.style={nodes={TRRSSE}}}
   \tikzset{row 1 column 6/.style={nodes={TRRSSE}}}
   \matrix (FIFO) [matrix of nodes,
                   nodes in empty cells,
                   nodes={draw,
                          ultra thin,
                          anchor=south,
                          rectangle,
                          text width=1.5em,
                          minimum height=3ex,
                         },
                   ] {
     &&&&&&&&\\
    };
    \begin{scope}[|-|,shorten >=1pt,shorten <=1pt,inner sep=1pt]
     \begin{scope}[opacity=.5]
      \draw[|-,densely dotted,shorten >=0pt] ($(FIFO-1-1.north west)+(0,4ex)$) -- ($(FIFO-1-1.north east)+(0,4ex)$);
      \draw ($(FIFO-1-2.north west)+(0,4ex)$) -- node[above,font=\footnotesize,midway] {MLSE states}   ($(FIFO-1-6.north east)+(0,4ex)$);
      \draw ($(FIFO-1-7.north west)+(0,4ex)$) -- node[above,font=\footnotesize,midway] {path register} ($(FIFO-1-9.north east)+(0,4ex)$);
     \end{scope}
     \draw[|-,densely dotted,shorten >=0pt] ($(FIFO-1-1.north west)+(0,1ex)$) -- ($(FIFO-1-1.north east)+(0,1ex)$);
     \draw ($(FIFO-1-2.north west)+(0,1ex)$) -- node[above,font=\footnotesize,midway] {RSSE states} ($(FIFO-1-4.north east)+(0,1ex)$);
     \draw ($(FIFO-1-5.north west)+(0,1ex)$) -- node[above,font=\footnotesize,midway] {path register} ($(FIFO-1-9.north east)+(0,1ex)$);
     \draw ($(FIFO-1-7.south west)-(0,1ex)$) --
     node[below=.5ex,font=\footnotesize,midway] (feedbacksrc) {\parbox{5em}{\centering enable branch}} ($(FIFO-1-4.south east)-(0,1ex)$);
     \draw ($(FIFO-1-2.south west)-(0,1ex)$) -- node[below=.5ex,font=\footnotesize,midway] {\parbox{3em}{\centering eventually\\punctured\\bit}} ($(FIFO-1-1.south west)-(0,1ex)$);
    \end{scope}
    \draw[->] (FIFO-1-9.south) ++(0,-1.5ex) node[right] {$t$} -- ++(-1em,0);
    \begin{scope}[line width=.1pt]
     \draw (FIFO-1-1.west) to[bend left=-15] ++(160:1cm) circle ( 1pt);
     \draw (FIFO-1-1.west) to[bend left=+15] ++(160:1cm) circle ( 1pt);
     \draw (FIFO-1-1.west) to[bend left=-5 ] ++(160:1cm) circle ( 1pt);
     \draw (FIFO-1-1.west) to[bend left=+5 ] ++(160:1cm) circle ( 1pt);
     \draw (FIFO-1-1.west) to[bend left=-15] ++(200:1cm) circle ( 1pt);
     \draw (FIFO-1-1.west) to[bend left=+15] ++(200:1cm) circle ( 1pt);
     \draw (FIFO-1-1.west) to[bend left=-5 ] ++(200:1cm) circle ( 1pt);
     \draw (FIFO-1-1.west) to[bend left=+5 ] ++(200:1cm) circle ( 1pt);
     \draw[fill] (FIFO-1-1.west) circle ( 1pt);
    \end{scope}
    \draw[densely dotted,->,shorten >=3ex] (feedbacksrc.south) -| ++ (0,-.3cm) -| ($(FIFO-1-1.west)+(-0.5cm,0)$);
  \end{tikzpicture}
 \end{center}\vspace*{-3ex}
 \caption{Graphical representation of our state design for punctured
          convolutional coding with and without reduced-state sequence estimation.}
 \label{fig:TrellisStateRSSE}
 \vspace*{-2ex}
\end{figure}
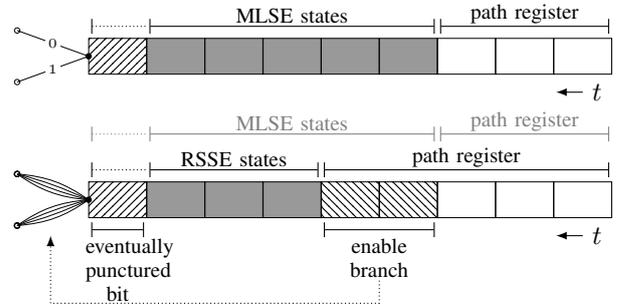

In Fig.~\ref{fig:TrellisStateRSSE} a single trellis state in the VA is
depicted as a FIFO. Input values to the FSM are represented by the branches at
the left-hand side, whereas values that drop of the FIFO are stored within the
state-specific path register $p_\mathrm{reg}(k,\lvec{s})$ at the right-hand
side.
During decoding when entering a trellis segment that is split, \ie, has an
increased number of states, the two possibilities for the punctured bit are
tracked using an additional delay element, indicated by the hatched block
\mbox{(\!\!\tikz[baseline=0ex]\draw[EXTEND] rectangle (1.5ex,1.5ex);)}.

When DFSE partitioning is performed, the states can be reduced as shown in
Fig.~\ref{fig:TrellisStateRSSE}. There, fewer FIFO elements are used to define
the trellis state, while the remaining ones are used as feedback to enable
branches for the next trellis step for that particular state.

An implementation of this algorithm needs to ensure that when entering a split
trellis segment, \ie, increased number of states, the right path register is
chosen as source for the decision feedback.
 
\subsection{State Partitioning                                } 
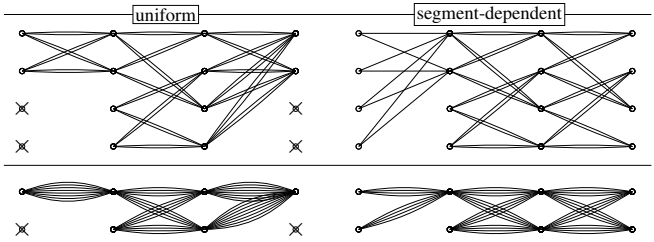
\begin{figure}
 \begin{tikzpicture}[x=12mm,y=5mm,line width=.1pt]
  \draw (0.8,-.5) --
        (7.9,-.5);
  \draw (0.8,3.5) --
          node[draw,pos=.25,anchor=center,fill=white,inner sep=1pt,font=\scriptsize] {uniform}
          node[draw,pos=.75,anchor=center,fill=white,inner sep=1pt,font=\scriptsize] {segment-dependent}
        (7.9,3.5);
  \begin{scope}
   \draw (1,0) circle(1pt) node {\textbf{$\times$}};
   \draw (1,1) circle(1pt) node {\textbf{$\times$}};
   \draw (4,0) circle(1pt) node {\textbf{$\times$}};
   \draw (4,1) circle(1pt) node {\textbf{$\times$}};
   \draw (1,3) circle(1pt) to[bend left=3]  (2,3) circle (1pt);
   \draw (1,2) circle(1pt) to[bend left=3]  (2,3) circle (1pt);
   \draw (1,3) circle(1pt) to[bend left=3]  (2,2) circle (1pt);
   \draw (1,2) circle(1pt) to[bend left=3]  (2,2) circle (1pt);
   \draw (1,3) circle(1pt) to[bend left=-3]  (2,3) circle (1pt);
   \draw (1,2) circle(1pt) to[bend left=-3]  (2,3) circle (1pt);
   \draw (1,3) circle(1pt) to[bend left=-3]  (2,2) circle (1pt);
   \draw (1,2) circle(1pt) to[bend left=-3]  (2,2) circle (1pt);
   \draw (2,3) circle(1pt) to[bend left=3]  (3,3) circle (1pt);
   \draw (2,2) circle(1pt) to[bend left=3]  (3,3) circle (1pt);
   \draw (2,1) circle(1pt) to[bend left=3]  (3,2) circle (1pt);
   \draw (2,0) circle(1pt) to[bend left=3]  (3,2) circle (1pt);
   \draw (2,3) circle(1pt) to[bend left=3]  (3,1) circle (1pt);
   \draw (2,2) circle(1pt) to[bend left=3]  (3,1) circle (1pt);
   \draw (2,1) circle(1pt) to[bend left=3]  (3,0) circle (1pt);
   \draw (2,0) circle(1pt) to[bend left=3]  (3,0) circle (1pt);
   \draw (2,3) circle(1pt) to[bend left=-3]  (3,3) circle (1pt);
   \draw (2,2) circle(1pt) to[bend left=-3]  (3,3) circle (1pt);
   \draw (2,1) circle(1pt) to[bend left=-3]  (3,2) circle (1pt);
   \draw (2,0) circle(1pt) to[bend left=-3]  (3,2) circle (1pt);
   \draw (2,3) circle(1pt) to[bend left=-3]  (3,1) circle (1pt);
   \draw (2,2) circle(1pt) to[bend left=-3]  (3,1) circle (1pt);
   \draw (2,1) circle(1pt) to[bend left=-3]  (3,0) circle (1pt);
   \draw (2,0) circle(1pt) to[bend left=-3]  (3,0) circle (1pt);
   \draw (3,3) circle(1pt) to[bend left=3]  (4,3) circle (1pt);
   \draw (3,2) circle(1pt) to[bend left=3]  (4,3) circle (1pt);
   \draw (3,1) circle(1pt) to[bend left=3]  (4,3) circle (1pt);
   \draw (3,0) circle(1pt) to[bend left=3]  (4,3) circle (1pt);
   \draw (3,3) circle(1pt) to[bend left=3]  (4,2) circle (1pt);
   \draw (3,2) circle(1pt) to[bend left=3]  (4,2) circle (1pt);
   \draw (3,1) circle(1pt) to[bend left=3]  (4,2) circle (1pt);
   \draw (3,0) circle(1pt) to[bend left=3]  (4,2) circle (1pt);
   \draw (3,3) circle(1pt) to[bend left=-3]  (4,3) circle (1pt);
   \draw (3,2) circle(1pt) to[bend left=-3]  (4,3) circle (1pt);
   \draw (3,1) circle(1pt) to[bend left=-3]  (4,3) circle (1pt);
   \draw (3,0) circle(1pt) to[bend left=-3]  (4,3) circle (1pt);
   \draw (3,3) circle(1pt) to[bend left=-3]  (4,2) circle (1pt);
   \draw (3,2) circle(1pt) to[bend left=-3]  (4,2) circle (1pt);
   \draw (3,1) circle(1pt) to[bend left=-3]  (4,2) circle (1pt);
   \draw (3,0) circle(1pt) to[bend left=-3]  (4,2) circle (1pt);
  \end{scope}
  \begin{scope}[xshift=.5\linewidth]
   \draw (1,3) circle(1pt) to[bend left=0]  (2,3) circle (1pt);
   \draw (1,2) circle(1pt) to[bend left=0]  (2,3) circle (1pt);
   \draw (1,1) circle(1pt) to[bend left=0]  (2,3) circle (1pt);
   \draw (1,0) circle(1pt) to[bend left=0]  (2,3) circle (1pt);
   \draw (1,3) circle(1pt) to[bend left=0]  (2,2) circle (1pt);
   \draw (1,2) circle(1pt) to[bend left=0]  (2,2) circle (1pt);
   \draw (1,1) circle(1pt) to[bend left=0]  (2,2) circle (1pt);
   \draw (1,0) circle(1pt) to[bend left=0]  (2,2) circle (1pt);
   \draw (2,3) circle(1pt) to[bend left=3]  (3,3) circle (1pt);
   \draw (2,2) circle(1pt) to[bend left=3]  (3,3) circle (1pt);
   \draw (2,1) circle(1pt) to[bend left=3]  (3,2) circle (1pt);
   \draw (2,0) circle(1pt) to[bend left=3]  (3,2) circle (1pt);
   \draw (2,3) circle(1pt) to[bend left=3]  (3,1) circle (1pt);
   \draw (2,2) circle(1pt) to[bend left=3]  (3,1) circle (1pt);
   \draw (2,1) circle(1pt) to[bend left=3]  (3,0) circle (1pt);
   \draw (2,0) circle(1pt) to[bend left=3]  (3,0) circle (1pt);
   \draw (2,3) circle(1pt) to[bend left=-3]  (3,3) circle (1pt);
   \draw (2,2) circle(1pt) to[bend left=-3]  (3,3) circle (1pt);
   \draw (2,1) circle(1pt) to[bend left=-3]  (3,2) circle (1pt);
   \draw (2,0) circle(1pt) to[bend left=-3]  (3,2) circle (1pt);
   \draw (2,3) circle(1pt) to[bend left=-3]  (3,1) circle (1pt);
   \draw (2,2) circle(1pt) to[bend left=-3]  (3,1) circle (1pt);
   \draw (2,1) circle(1pt) to[bend left=-3]  (3,0) circle (1pt);
   \draw (2,0) circle(1pt) to[bend left=-3]  (3,0) circle (1pt);
   \draw (3,3) circle(1pt) to[bend left=3]  (4,3) circle (1pt);
   \draw (3,2) circle(1pt) to[bend left=3]  (4,3) circle (1pt);
   \draw (3,1) circle(1pt) to[bend left=3]  (4,2) circle (1pt);
   \draw (3,0) circle(1pt) to[bend left=3]  (4,2) circle (1pt);
   \draw (3,3) circle(1pt) to[bend left=3]  (4,1) circle (1pt);
   \draw (3,2) circle(1pt) to[bend left=3]  (4,1) circle (1pt);
   \draw (3,1) circle(1pt) to[bend left=3]  (4,0) circle (1pt);
   \draw (3,0) circle(1pt) to[bend left=3]  (4,0) circle (1pt);
   \draw (3,3) circle(1pt) to[bend left=-3]  (4,3) circle (1pt);
   \draw (3,2) circle(1pt) to[bend left=-3]  (4,3) circle (1pt);
   \draw (3,1) circle(1pt) to[bend left=-3]  (4,2) circle (1pt);
   \draw (3,0) circle(1pt) to[bend left=-3]  (4,2) circle (1pt);
   \draw (3,3) circle(1pt) to[bend left=-3]  (4,1) circle (1pt);
   \draw (3,2) circle(1pt) to[bend left=-3]  (4,1) circle (1pt);
   \draw (3,1) circle(1pt) to[bend left=-3]  (4,0) circle (1pt);
   \draw (3,0) circle(1pt) to[bend left=-3]  (4,0) circle (1pt);
  \end{scope}
  \begin{scope}[yshift=-1.1cm]
   \draw (1,0) circle(1pt) node {\textbf{$\times$}};
   \draw (4,0) circle(1pt) node {\textbf{$\times$}};
   \draw (1,1) circle(1pt) to[bend left=21]  (2,1) circle (1pt);
   \draw (1,1) circle(1pt) to[bend left=15]  (2,1) circle (1pt);
   \draw (1,1) circle(1pt) to[bend left=9]  (2,1) circle (1pt);
   \draw (1,1) circle(1pt) to[bend left=3]  (2,1) circle (1pt);
   \draw (1,1) circle(1pt) to[bend left=-3]  (2,1) circle (1pt);
   \draw (1,1) circle(1pt) to[bend left=-9]  (2,1) circle (1pt);
   \draw (1,1) circle(1pt) to[bend left=-15]  (2,1) circle (1pt);
   \draw (1,1) circle(1pt) to[bend left=-21]  (2,1) circle (1pt);
   \draw (2,1) circle(1pt) to[bend left=9]  (3,1) circle (1pt);
   \draw (2,0) circle(1pt) to[bend left=9]  (3,1) circle (1pt);
   \draw (2,1) circle(1pt) to[bend left=9]  (3,0) circle (1pt);
   \draw (2,0) circle(1pt) to[bend left=9]  (3,0) circle (1pt);
   \draw (2,1) circle(1pt) to[bend left=3]  (3,1) circle (1pt);
   \draw (2,0) circle(1pt) to[bend left=3]  (3,1) circle (1pt);
   \draw (2,1) circle(1pt) to[bend left=3]  (3,0) circle (1pt);
   \draw (2,0) circle(1pt) to[bend left=3]  (3,0) circle (1pt);
   \draw (2,1) circle(1pt) to[bend left=-3]  (3,1) circle (1pt);
   \draw (2,0) circle(1pt) to[bend left=-3]  (3,1) circle (1pt);
   \draw (2,1) circle(1pt) to[bend left=-3]  (3,0) circle (1pt);
   \draw (2,0) circle(1pt) to[bend left=-3]  (3,0) circle (1pt);
   \draw (2,1) circle(1pt) to[bend left=-9]  (3,1) circle (1pt);
   \draw (2,0) circle(1pt) to[bend left=-9]  (3,1) circle (1pt);
   \draw (2,1) circle(1pt) to[bend left=-9]  (3,0) circle (1pt);
   \draw (2,0) circle(1pt) to[bend left=-9]  (3,0) circle (1pt);
   \draw (3,1) circle(1pt) to[bend left=21]  (4,1) circle (1pt);
   \draw (3,0) circle(1pt) to[bend left=21]  (4,1) circle (1pt);
   \draw (3,1) circle(1pt) to[bend left=15]  (4,1) circle (1pt);
   \draw (3,0) circle(1pt) to[bend left=15]  (4,1) circle (1pt);
   \draw (3,1) circle(1pt) to[bend left=9]  (4,1) circle (1pt);
   \draw (3,0) circle(1pt) to[bend left=9]  (4,1) circle (1pt);
   \draw (3,1) circle(1pt) to[bend left=3]  (4,1) circle (1pt);
   \draw (3,0) circle(1pt) to[bend left=3]  (4,1) circle (1pt);
   \draw (3,1) circle(1pt) to[bend left=-3]  (4,1) circle (1pt);
   \draw (3,0) circle(1pt) to[bend left=-3]  (4,1) circle (1pt);
   \draw (3,1) circle(1pt) to[bend left=-9]  (4,1) circle (1pt);
   \draw (3,0) circle(1pt) to[bend left=-9]  (4,1) circle (1pt);
   \draw (3,1) circle(1pt) to[bend left=-15]  (4,1) circle (1pt);
   \draw (3,0) circle(1pt) to[bend left=-15]  (4,1) circle (1pt);
   \draw (3,1) circle(1pt) to[bend left=-21]  (4,1) circle (1pt);
   \draw (3,0) circle(1pt) to[bend left=-21]  (4,1) circle (1pt);
  \end{scope}
  \begin{scope}[yshift=-1.1cm,xshift=.5\linewidth]
   \draw (1,1) circle(1pt) to[bend left=9]  (2,1) circle (1pt);
   \draw (1,0) circle(1pt) to[bend left=9]  (2,1) circle (1pt);
   \draw (1,1) circle(1pt) to[bend left=3]  (2,1) circle (1pt);
   \draw (1,0) circle(1pt) to[bend left=3]  (2,1) circle (1pt);
   \draw (1,1) circle(1pt) to[bend left=-3]  (2,1) circle (1pt);
   \draw (1,0) circle(1pt) to[bend left=-3]  (2,1) circle (1pt);
   \draw (1,1) circle(1pt) to[bend left=-9]  (2,1) circle (1pt);
   \draw (1,0) circle(1pt) to[bend left=-9]  (2,1) circle (1pt);
   \draw (2,1) circle(1pt) to[bend left=9]  (3,1) circle (1pt);
   \draw (2,0) circle(1pt) to[bend left=9]  (3,1) circle (1pt);
   \draw (2,1) circle(1pt) to[bend left=9]  (3,0) circle (1pt);
   \draw (2,0) circle(1pt) to[bend left=9]  (3,0) circle (1pt);
   \draw (2,1) circle(1pt) to[bend left=3]  (3,1) circle (1pt);
   \draw (2,0) circle(1pt) to[bend left=3]  (3,1) circle (1pt);
   \draw (2,1) circle(1pt) to[bend left=3]  (3,0) circle (1pt);
   \draw (2,0) circle(1pt) to[bend left=3]  (3,0) circle (1pt);
   \draw (2,1) circle(1pt) to[bend left=-3]  (3,1) circle (1pt);
   \draw (2,0) circle(1pt) to[bend left=-3]  (3,1) circle (1pt);
   \draw (2,1) circle(1pt) to[bend left=-3]  (3,0) circle (1pt);
   \draw (2,0) circle(1pt) to[bend left=-3]  (3,0) circle (1pt);
   \draw (2,1) circle(1pt) to[bend left=-9]  (3,1) circle (1pt);
   \draw (2,0) circle(1pt) to[bend left=-9]  (3,1) circle (1pt);
   \draw (2,1) circle(1pt) to[bend left=-9]  (3,0) circle (1pt);
   \draw (2,0) circle(1pt) to[bend left=-9]  (3,0) circle (1pt);
   \draw (3,1) circle(1pt) to[bend left=9]  (4,1) circle (1pt);
   \draw (3,0) circle(1pt) to[bend left=9]  (4,1) circle (1pt);
   \draw (3,1) circle(1pt) to[bend left=9]  (4,0) circle (1pt);
   \draw (3,0) circle(1pt) to[bend left=9]  (4,0) circle (1pt);
   \draw (3,1) circle(1pt) to[bend left=3]  (4,1) circle (1pt);
   \draw (3,0) circle(1pt) to[bend left=3]  (4,1) circle (1pt);
   \draw (3,1) circle(1pt) to[bend left=3]  (4,0) circle (1pt);
   \draw (3,0) circle(1pt) to[bend left=3]  (4,0) circle (1pt);
   \draw (3,1) circle(1pt) to[bend left=-3]  (4,1) circle (1pt);
   \draw (3,0) circle(1pt) to[bend left=-3]  (4,1) circle (1pt);
   \draw (3,1) circle(1pt) to[bend left=-3]  (4,0) circle (1pt);
   \draw (3,0) circle(1pt) to[bend left=-3]  (4,0) circle (1pt);
   \draw (3,1) circle(1pt) to[bend left=-9]  (4,1) circle (1pt);
   \draw (3,0) circle(1pt) to[bend left=-9]  (4,1) circle (1pt);
   \draw (3,1) circle(1pt) to[bend left=-9]  (4,0) circle (1pt);
   \draw (3,0) circle(1pt) to[bend left=-9]  (4,0) circle (1pt);
  \end{scope}
 \end{tikzpicture}\vspace*{-3ex}
 \caption{Illustration of uniform and segment-dependent partitioning for
          \mbox{$J=\left\{ 1,2 \right\}$}.}
 \label{fig:trellispart}
 \vspace*{-2ex}
\end{figure}

As already mentioned, a DFSE partitioning of the first order, \ie, reducing the
number of states in each trellis segment by a factor of two, combining states
that differ in the eldest position, into hyperstates. Hence, when applied to a
punctured TCM, each trellis segment undergoes the same partitioning. A
resulting reduced-state trellis is depicted on the left-hand side of
Fig.~\ref{fig:trellispart} for $J=1$ and $J=2$.
As a consequence, non-existing states from the original trellis are also
partitioned (\cf, first trellis segment).
 
Due to the reduction of non-existing states, and hence reduced minimum
Euclidean distance, we propose to partition those segments first that are split
and keeping the others unpartitioned. As can be seen from the first-level
state reduction in Fig.~\ref{fig:trellispart} (right-hand side, above), the
first segment contains two transitions in each state, and thus is not
partitioned, whereas in segment two and three, each state has four transitions,
due to the state partitioning.

Apparently, the first segment may be handled with a full state VA, while
the other two segments need to be decoded via RSSE using the path register in
each state.

The right-hand side of Fig.~\ref{fig:trellispart} show the segment-depending
partitioning and state reduction for $J=1$ and $J=2$. The segments for $J=1$
show four, and eight transitions per state, respectively. At this point RSSE
has to consider a different amount of information in the path register for each
segment.

The resulting trellis shows a less decreased minimum Euclidean distance when
compared to the uniform partitioning but has a slightly higher computational
complexity because of the extra states. Thus, the segment-dependent
partitioning technique enables an even more flexible way to trade between
complexity and performance.
 
\section{Numerical Results                                    } \label{sec:results}
In this section we will give numerical simulation results and investigate
several ISI-channels for \emph{punctured} TCM. We analyse the decoder
complexity as \emph{number of branch metric calculations} per \emph{information
bit} and show that this approach enables a flexible trade-off between
computational complexity and performance.

Due to the minimum phase characteristics of the ISI-channel the partitioning of
the trellis states into hyperstates leads to the smallest possible loss in
Euclidean distance. Hence, if, for instance, the ISI-channel is an equal tab
delay line, the loss in Euclidean distance is higher than for an exponentially
decaying channel because of the premature decisions for a surviving path.

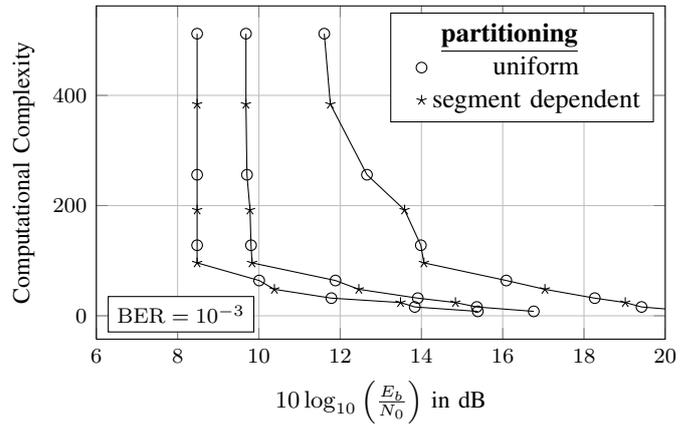
\begin{figure}
 \begin{center}
  \begin{tikzpicture}
   \begin{axis}[
                width=.5\textwidth,
                height=6cm,
                xlabel={$10\log_{10}\left(\frac{E_b}{N_0}\right)$ in dB},
                ylabel={Computational Complexity},
                cycle list={
                            every mark/.append style={fill=black},mark=o\\%
                            every mark/.append style={fill=black},mark=star\\%
                           },
                grid=both,
                xmin=6,xmax=20,
               ]
    \addlegendimage{empty legend} 
    \addplot+[only marks] table[x index=0,y index=1] {JOURNAL-Channel-ASK-complexity-1-1.csv};
    \addplot+[only marks] table[x index=0,y index=1] {JOURNAL-Channel-ASK-complexity-1-2.csv};
    \addplot+[only marks] table[x index=0,y index=1] {JOURNAL-Channel-ASK-complexity-2-1.csv};
    \addplot+[only marks] table[x index=0,y index=1] {JOURNAL-Channel-ASK-complexity-2-2.csv};
    \addplot+[only marks] table[x index=0,y index=1] {JOURNAL-Channel-ASK-complexity-3-1.csv};
    \addplot+[only marks] table[x index=0,y index=1] {JOURNAL-Channel-ASK-complexity-3-2.csv};
    \addplot[black] table[x index=0,y index=1] {JOURNAL-Channel-ASK-complexity-rsse-1.csv}
       node[pos=0,pin={[pin distance=.5em,rotate=90,inner sep=1pt,fill=white]175:{\scriptsize{exponential channel}}}] {};
    \addplot[black] table[x index=0,y index=1] {JOURNAL-Channel-ASK-complexity-rsse-2.csv}
       node[pos=0,pin={[pin distance=.5em,rotate=90,inner sep=1pt,fill=white]175:{\scriptsize{linearly decreasing channel}}}] {};
    \addplot[black] table[x index=0,y index=1] {JOURNAL-Channel-ASK-complexity-rsse-3.csv}
       node[pos=0,pin={[pin distance=.5em,rotate=90,inner sep=1pt,fill=white]175:{\scriptsize{equal channel}}}] {};;
    \addlegendentry{\hspace{-.7cm}\underline{\textbf{partitioning}}}
    \addlegendentry{uniform}
    \addlegendentry{segment dependent}
    \node[draw,fill=white,anchor=south west,font={\footnotesize}] at (rel axis cs:0.02,0.02) {$\mathrm{BER} = 10^{-3}$};
   \end{axis}
  \end{tikzpicture}\vspace*{-4ex}
 \end{center}
 \caption{Decoding complexity for a P-TCM transmission scheme with generator
          polynomials $\left[ 13\, 15 \right]_\mathrm{oct}$, puncturing scheme
          $\left[\,(1\,1)^\top\,(0\,1)^\top\,\right]$ \mbox{(Rate: $\frac43$)} natural
          labeling and $4$-ASK signaling over three different ISI-channels
          $h_\mathrm{exp}[\kappa]$ ($\kappa_0=1$), $h_\mathrm{lin}[\kappa]$,
          $h_\mathrm{equ}[\kappa]$ (solid: Uniform partitioning, dashed:
          Segment-dependent partitioning).}
 \label{fig:complexity-Channel-ASK}
 \vspace*{-2ex}
\end{figure}

To see this effect we conducted simulations over different ISI-channels of unit
energy and plotted the complexity number over the required
$\frac{E_\mathrm{b}}{N_0}$ to achieve a bit error probability of less than
$10^{-3}$.
The unit energy channels are defined as:
\begin{align*}
h_\mathrm{exp}[\kappa] & = \frac{1}{\sqrt{\sum\limits_\kappa{|h_\mathrm{exp}[\kappa]|^2}}}\,\e^{(-\kappa/\kappa_0)} & \text{for } 0\leq\kappa\leq L\\
h_\mathrm{lin}[\kappa] & = \frac{1}{\sqrt{\sum\limits_\kappa{|h_\mathrm{lin}[\kappa]|^2}}}\,\frac{L-\kappa+1}{L+1} & \text{for } 0\leq\kappa\leq L\\
h_\mathrm{equ}[\kappa] & = \frac{1}{\sqrt{\sum\limits_\kappa{|h_\mathrm{equ}[\kappa]|^2}}} = \frac{1}{\sqrt{L}}    & \text{for } 0\leq\kappa\leq L
\end{align*}
The results can be seen in Fig.~\ref{fig:complexity-Channel-ASK}. As should be
clear to the reader, the loss in Euclidean distance is smallest for an
exponentially decaying ISI-channel.

\section{Conclusion                                           } \label{sec:conclusion}

It has been shown that TCM can be extended by puncturing. Furthermore, an
efficient MLSE decoder for P-TCM over ISI-channels was proposed and
investigated.

The numerical simulation results clearly show that we can achieve a soft
trade-off between spectral and power efficiency easier and more flexibly than
by means of traditional TCM.
 
\vspace*{-4ex}
%
%
\IEEEtriggeratref{1}
\IEEEtriggercmd{\enlargethispage{1in}}
\bibliographystyle{IEEEtran}
\bibliography{IEEEabrv,main,mine}
\end{document}